\documentclass[11pt]{article}
\usepackage {amsfonts}
\usepackage {graphicx}
\usepackage {epsfig}
\usepackage{amssymb}
\usepackage{amsmath}
\usepackage {longtable}
\usepackage[english]{babel}
\usepackage{cite}

\begin{document}

\title{About possible influence of birefringence effect on processes of production
(photoproduction, electroproduction) of vector mesons (particles
with the spin $S \ge 1$) in nuclei}

\author{V.G. Baryshevsky \\ Research Institute for Nuclear Problems, Belarusian State
University,\\ 11 Bobruyskaya Str., Minsk 220050, Belarus,
\\ e-mail: bar@inp.minsk.by}

\maketitle

\begin{center}
\begin{abstract}
It is shown that birefringence effect influences on production  of
particles with the spin $S \ge 1$ at collisions of high energy
particles.
\end{abstract}
\end{center}

\section{INTRODUCTION}

Collision of high energy particles (proton, electron,
$\gamma$-quanta, nucleus) with nucleus yields a lot of hadronic
processes inside nuclei, which are accompanied by appearance of
secondary particles (vector mesons, $\Omega^-$ hyperons and so
on).
In particular the processes of photoproduction (electroproduction)
of vector mesons by nuclei have been studying since the
photoproduction vertex for hadronic probes inside nuclei is well
known and the analysis of the results is simple and more reliable.

Moreover, experiments demonstrate production of both
longitudinally (L) and transversally (T) polarized vector mesons
and L/T-ratios depend on $Q^2$ \cite{dorohov}.

According to \cite{01} consideration of photoproduction of vector
mesons inside nuclei and their rescattering via strong
interactions within Glauber multiple scattering theory allows to
consider many peculiarities of the process important for
understanding unconventional effects, like color transparency.
In \cite{01} (as well as in others) theoretical consideration of
processes of rescattering of produced particles inside the nucleus
is done without taking into account possible influence of produced
particle spin on rescattering inside the nucleus.
For the first glance there is no necessity to consider spin
effects, caused by rescattering inside the nucleus, because
usually nuclei are nonpolarized.
However, this is a hasty conclusion.

In \cite{2,3,4} it was shown that birefringence phenomenon
appears when a deuteron (or a particle with spin $S \ge 1$) passes
through homogenies and isotropic matter.
Birefringence phenomenon is the effect of spin rotation around
momentum direction and spin dichroism (i.e. dependence of
absorbtion coefficient on spin direction with respect to deuteron
momentum).
These effects are aroused due to intrinsic anisotropy, which is
inherent for particles with spin $S \ge 1$.
In particular this is well known for deuterons \cite{5} and
$\Omega^-$-hyperons \cite{Ger}, for which the ground state is the
mix of $S$ and $D$ waves.

For a particle, produced inside a nucleus and moving though the
nuclear matter, the conception of refraction index can be applied,
too.

%

Birefringence effect  can be described by the spin-dependent index
of refraction:
\begin{equation}
\hat{n}=1+\frac{2 \pi \rho}{k^2} \hat{f} (0),
 \label{n}
\end{equation}
where $\rho$ is the density of scatterers in matter (the number of
scatterers in 1 cm$^3$), $k$ is the particle wavenumber,
$\hat{f}(0)$ is used to denote the amplitude of zero-angle elastic
coherent scattering of a particle by a scattering center, this
amplitude is an operator acting in the particle spin space.

Refraction of particles in matter (either conventional or nuclear)
implies existence of optical pseudopotential \cite{2,3}:
\begin{equation}
\hat{V}_{eff}= - \frac{2 \pi \hbar^2}{m \gamma} \hat{f}(0),
 \label{Vopt}
\end{equation}
where m particle mass and $\gamma$ is its Lorentz factor.

For a particle with the spin 1 (deuteron, vector mesons) the
amplitude of zero-angle elastic coherent scattering by a
nonpolarized scatterer can be expressed in the following general
form:
\begin{equation}
\hat{f} (0)=d+d_1 (\vec{S} \vec{n})^2,
 \label{f}
\end{equation}
where $\vec{S}$ is the operator of particle spin,
$\vec{n}=\frac{\vec{k}}{k}$ is the unit vector along particle
momentum.
 The angle of spin rotation is determined by
$Re~d_1$, while $Im~d_1$ describes dichroism.
Therefore, study of birefringence effect allows to find the
spin-dependent part $d_1$ of the amplitude of zero-angle elastic
coherent scattering.

If scatterers are polarized, the expression for the scattering
amplitude includes also terms depending on the scatterer
polarization \cite{2,3}.
In particular, for $\gamma$-quanta passing through matter with
polarized nuclei the amplitude of zero-angle scattering can be
expressed as follows:
\begin{equation}
\hat{f}_{\gamma N} (0)=A (\vec{e}^{\prime *} \vec{e}) + i G
 [\vec{e}^{\prime *}  \times \vec{e}]\vec{P} + B {e}^{\prime *}_i
{e}_k {Q}_{ik} + D {n}_{\gamma i} n_{\gamma k} {Q}_{ik},
 \label{fpol}
\end{equation}
where $\vec{e}$ is the vector of incident $\gamma$-quanta
polarization, $\vec{e}^{\prime}$ is  the vector of scattered
$\gamma$-quanta polarization, $^*$ means complex conjugation,
$\vec{P}$ is the scatterer polarization vector, ${Q}_{ik}$ is the
rank two polarization tensor (quadrupolarization tensor) of
scatterer, $\vec{n}_{\gamma}$ is the unit vector along the
$\gamma$-quantum momentum.

The term containing $G$ ($G\vec{P}$ is the gyration vector)
describes effects of spin rotation and circular dichroism in the
target with polarized nuclei (nucleons) \cite{8}, the term
proportional $\sim B$ describes $\gamma$-quanta birefringence in
polarized target \cite{8}, the term containing $\vec{n}_{\gamma}$
describes dependence of $\gamma$-quanta absorbtion on the
orientation of quadrupolarization tensor ${Q}_{ik}$ of scatterer
(similarly the term $(\vec{S} \vec{n})^2$ in (\ref{f})).

Now spin dichroism was experimentally observed for deuterons with
the energy 10-20 MeV \cite{ex1,ex2} passing through a carbon
target.
In these experiments the value $Im~d_1$ was found for the first
time i.e. the difference in the deuteron scattering cross-sections
$\Delta \sigma = \sigma_{m=\pm1} - \sigma_{m=0}$, where
$\sigma_{m=\pm 1}$ is the total scattering cross-section for a
deuteron with the spin projection onto $\vec{n}$ is $m=\pm 1$ and
$\sigma_{m=0}$ is the total scattering cross-section for a
deuteron with the spin projection onto $\vec{n}$ is $m=0$.
Tensor polarization of deuteron beam with the energy 5.5 GeV
travelling through matter was observed in recent experiments
\cite{ex3}, too.

It should be mentioned that $\Delta \sigma \ne 0$ means that spin
features of a vector meson produced inside a nucleus (another
particle with the spin  $\ge 1$) will differ from spin properties
of the particle produced by a stand-alone nucleon (which does not
compose the nucleus)
%
%
that results, in particular, in change of L/T-ratio.

To describe rescattering processes in the energy range, where $Re
f(0)<< Im f(0)$, the expressions obtained in \cite{01} can be
used.
But the total cross-sections of vector meson production there
should be replaced by $\sigma_{M=\pm 1}$ or $\sigma_{M=0}$.

Additional analysis is necessary for $Re f(0)$ comparable or
larger than $Im f(0)$; it will done separately.


It should be mentioned that in general case  two correlations
present in photoproduction: correlation $[\vec{e}^{*} \vec{e}]
\vec{J}$ (where $\vec{J}$ is the operator of produced particle
spin) is sensitive to circular polarization of photons and
produced particle has vector polarization.
Correlation $(\vec{e} \vec{J})^2$, which is sensitive to linear
polarization of photon, corresponds production of particle with
tensor polarization.
Due to birefringence effect produced particles are absorbed by the
nucleus differently.
Therefore,  yield of vector-mesons  depends on photon polarization
i.e. production cross-section are different for different
polarization of incident photons $\sigma_{circ} \ne \sigma_{lin}$.


In the present paper it is shown that  spin-orbital interaction
contributes to the birefringence effect for particles with the
spin $\ge 1$ along with central interaction.

Consideration is made by the example of  contribution from
spin-orbital interaction to birefringence effect, which is caused
by interaction of particle with the coulomb field of the nucleus.

\section{The amplitude of forward scattering for a particle with the spin $\ge 1$ in coulomb field}

\subsection{Spin dichroism caused by the spin-orbital electromagnetic interaction}

 Let us consider first scattering of a structureless
charged particle with the spin $\vec{S}$ by a coulomb center.
The energy of spin-orbital interaction of such a particle the
electric field can be expressed as follows:
\begin{equation}
\hat{V}_{em}=i b \vec{S} [ \vec{E} \times \vec{\nabla} ],
 \label{Vem}
\end{equation}
where $b=(g-2+\frac{2}{\gamma+1}) \mu_{B}
\frac{\hbar}{mc\gamma^3}$, nuclear Bohr magneton $\mu_{B}=\frac{e
\hbar}{2 m c}$, $g$ is the gyromagnetic ratio, for deuteron
$g=1.72$, $\gamma$ is the particle Lorentz factor.
Therefore, the amplitude of particle scattering by electric field
can be written in the first Born approximation as follows:
\begin{eqnarray}
& & \hat{f}(\vec{k}^{\prime} - \vec{k})= \hat{f}_c
(\vec{k}^{\prime} - \vec{k})-\frac{m \gamma}{2 \pi \hbar^2}
\hat{V}_{em} (\vec{k}^{\prime} - \vec{k}) = \hat{f}_c
(\vec{k}^{\prime} - \vec{k}) - \frac{m}{2 \pi \hbar^2}b~ \vec{S}
\,[ \vec{E}(\vec{k}^{\prime} - \vec{k}) \times  \vec{k}  ] =
\nonumber
\\
& & = \hat{f}_c (\vec{k}^{\prime} - \vec{k}) - \frac{m}{2 \pi
\hbar^2} b ~ \Phi (\vec{k}^{\prime} - \vec{k}) \, \vec{S} \, [
\vec{k}^{\prime} \times \vec{k} ] ,
 \label{f1}
\end{eqnarray}
where $\hat{f}_c (\vec{k}^{\prime} - \vec{k})$ is the amplitude of
coulomb scattering of a charge by charge, $\Phi (\vec{k}^{\prime}
- \vec{k})$ is the Fourier transform of the coulomb potential.
For a shielded coulomb potential $\Phi (\vec{k}^{\prime} -
\vec{k})= \frac{4 \pi Z e}{(\vec{k}^{\prime} - \vec{k})^2 +
\kappa^2} \rho (\vec{k}^{\prime} - \vec{k})$, $Z e$ is the
scattering center charge, $\kappa = \frac{1}{R_{sh}}$, $R_{sh}$ is
the shielding radius, $\rho (\vec{k}^{\prime} - \vec{k})= \int
e^{-i (\vec{k}^{\prime} - \vec{k}) \vec{r}} \rho (\vec{r}) d^3 r$
is the Fourier transform of the charge distribution density in the
scatterer $\rho (\vec{r})$.
When charge distribution is spherically symmetric $\rho
(\vec{k}^{\prime} - \vec{k})= \rho (|\vec{k}^{\prime} -
\vec{k}|)$.

From (\ref{f1}) follows that amplitude of forward scattering in
the first Born approximation is determined only by Coulomb
interaction of the charges of colliding particles, while
contribution from spin-orbital interaction is zero.

Let us obtain now contribution from spin-orbital interaction to
total scattering cross-section:
\begin{equation}
\sigma = \int f_{em}^{+}(\vec{k}^{\prime} - \vec{k})
f_{em}^{+}(\vec{k}^{\prime} - \vec{k}) d \Omega_{\vec{k}^{\prime}}
= \left( \frac{m \gamma}{2 \pi \hbar^2} \right)^2 b^2 \int \Phi^2
(\vec{k}^{\prime} - \vec{k}) (\vec{S} [\vec{k}^{\prime} \times
\vec{k}])^2 d \Omega_{\vec{k}^{\prime}},
 \label{sigma}
\end{equation}
i.e.
\begin{equation}
\sigma = \left( \frac{m \gamma}{2 \pi \hbar^2} \right)^2 b^2 \int
\Phi^2 (\vec{k}^{\prime} - \vec{k}) (\vec{S} [\vec{k}^{\prime}
\times \vec{k}])(\vec{S} [\vec{k}^{\prime} \times \vec{k}]) d
\Omega_{\vec{k}^{\prime}} = \left( \frac{m \gamma}{2 \pi \hbar^2}
\right)^2 b^2 \int \Phi^2 (\vec{k}^{\prime} - \vec{k})
(\vec{k}^{\prime}  [\vec{k} \times \vec{S}] ) (\vec{k}^{\prime}
[\vec{k} \times \vec{S}] ) d \Omega_{\vec{k}^{\prime}},
 \label{sigma1}
\end{equation}

Let us direct the axis $z$ along the wavevector $\vec{k}$.

In this case $\Phi^2 (\vec{k}^{\prime} - \vec{k}) = \Phi^2
(\sqrt{k^{\prime 2}_{\perp} + (k_z^{\prime}-k^2)^2})$ i.e.
$\Phi^2$ does not depend on the azimuth angle $\phi$ of vector
$\vec{k}$, $k^{\prime}_{\perp}$ is the component of
$\vec{k}^{\prime}$ perpendicular to $\vec{k}$.
Obviously $\Phi^2 (\vec{k}^{\prime} - \vec{k})=\Phi^2 (\sqrt{ 2
(1-cos \vartheta)})$, $\vartheta$ is the azimuth angle of
$\vec{k}^{\prime}$.
As a result, the expression in (\ref{sigma1}) can be integrated
over $\phi$ that gives:
\begin{equation}
\sigma = \left( \frac{m \gamma}{2 \pi \hbar^2} \right)^2 b^2 k^4
\pi
\int_0^{\pi} \Phi^2 (\sqrt{ 2 (1-cos \vartheta)})
sin^2 \vartheta (S (S+1) - S_z^2) sin \vartheta d \vartheta.
 \label{sigma3}
\end{equation}
From here it immediately follows that total cross-section
$\sigma_{M=\pm 1}$ for a particle with spin and magnetic quantum
number $M=\pm 1$ is not equal to total cross-section
$\sigma_{M=0}$ for a particle with spin having magnetic quantum
number $M=0$.

Difference $\Delta \sigma $ can be expressed as:
\begin{equation}
\Delta \sigma = \sigma_{M=\pm 1} - \sigma_{M=0}=
- \frac{m^2 \gamma^2}{4 \pi^2 \hbar^4}  b^2 k^4 \pi
\int_0^{\pi} \Phi^2 (\sqrt{ 2 (1-cos \vartheta)})
sin^3 \vartheta d \vartheta.
 \label{sigma4}
\end{equation}
Fourier transform of the Coulomb potential looks like:
\begin{equation}
\Phi (\vec{k}^{\prime} - \vec{k}) =
\frac{4 \pi Z e}{(\vec{k}^{\prime} - \vec{k})^2 + \kappa^2}=
\frac{4 \pi Z e}{2 {k}^{2}(1-cos \vartheta) + \kappa^2}
 \label{Coulomb}
\end{equation}
Therefore, for particle scattered by the point coulomb center
$\rho (\vec{k}^{\prime} - \vec{k})=1$
\begin{equation}
\Delta \sigma =
- \frac{m^2 \gamma^2}{4 \pi^2 \hbar^4}  b^2 k^4 \pi
\int_{-1}^{1}  \frac{(4 \pi Z e)^2 (1- \kappa^2)}{[2 k^2 (1-x)+
\kappa^2]} dx.
 \label{sigma5}
\end{equation}
Integration gives:
\begin{equation}
\Delta \sigma =
\frac{m^2 \gamma^2}{8 \pi \hbar^4}  b^2  (4 \pi Z e)^2
\left(2+\left( 1+ \frac{\kappa^2}{2 k^2} \right) ln
\frac{\kappa^2}{2 k^2 (1+ \frac{\kappa^2}{2 k^2})}\right)
 \label{sigma6}
\end{equation}

It is interesting that for deuteron, having $g=1.72$ at the energy
$E \approx 11.5$ GeV from (\ref{sigma6}) it follows that $\Delta
\sigma =0$.

\subsection{Particle spin rotation around the momentum due to spin-orbital electromagnetic interaction}

According to \cite{2} dispersion relations for the zero-angle
scattering amplitude determines $Re f_{\pm}(0) \ne Re f_0 (0)$.

As a result, spin-orbital interaction causes difference of
refraction indices $n_{\pm}-n_0 \ne 0$ and,therefore,
birefringence effect (i.e. the effect of spin rotation around the
direction of incident particle momentum) and spin dichroism.

To find the spin-dependent $Re f(0)$ let us use the second Born
approximation (note that according to the above in the first Born
approximation spin-orbital interaction does not contribute to
$\hat{f} (0)$).

The scattering amplitude in the second Born approximation reads:
\begin{equation}
\hat{f}^{(2)}(0)= \frac{m^2 \gamma^2}{8 \pi^4 \hbar^4} P
\int\frac{\hat{V}^{+} (\vec{k}- \vec{q}) V(\vec{q}-\vec{k})}{q^2 -
k^2} d^3q +
i \frac{m^2 \gamma^2}{8 \pi^3 \hbar^4} \int \hat{V}^{+} (\vec{k}-
\vec{q}) V(\vec{q}-\vec{k}) \delta (q^2 - k^2) d^3 q,
 \label{f2}
\end{equation}
where $P$ indicates principal integral value.
The imaginary part of amplitude $Im \hat{f}^{(2)} (0) = \frac{k}{4
\pi} \sigma$ and the expression for the cross-section $\sigma =
\frac{4 \pi}{k} Im   \hat{f}^{(2)} (0)$ certainly agrees with the
 expression (\ref{sigma}).

 Integration over the angles of $\vec{q}$ provides for $Re \hat{f}^{(2)
 (0)}$ the following expression:
\begin{eqnarray}
Re \hat{f}^{(2)}(0)= \frac{m^2 \gamma^2}{8 \pi^4 \hbar^4} P \int
\frac{b^2 \Phi^2 (\vec{q} - \vec{k})(\vec{S} [\vec{q} \times
\vec{k}])^2}{q^2 - k^2} q^2 dq d \Omega_q = \\
= \frac{m^2 \gamma^2}{8 \pi^3 \hbar^4} b^2 P \int \frac{q^4
k^2}{q^2-k^2} \Phi^2 (q \sqrt{2 (1-cos \vartheta)}) (S
(S+1)-S_z^2) sin ^3 \vartheta d \vartheta d q, \nonumber
 \label{Ref}
\end{eqnarray}
the axes $z$ is directed along the momentum of the incident
particle, i.e. along $\vec{k}$.

For particles with the spin 1 the real part of the difference in
amplitudes $f_{\pm 1} (0)$ and $f_0 (0)$ is expressed as:

\begin{equation}
 Re \Delta {f}(0)= Re {f}^{(2)}_{\pm 1}(0) -Re {f}^{(2)}_0(0)=
\frac{m^2 \gamma^2 b^2 k^2}{8 \pi^3 \hbar^4} P \int
\frac{q^4}{q^2-k^2} \Phi^2 (q \sqrt{2 (1-cos \vartheta)})  sin ^3
\vartheta d \vartheta d q,
 \label{DeltaRef}
\end{equation}

Let us find expression for $Re \Delta f$ for a particle scattered
by a point Coulomb scatterer:
\begin{equation}
 Re \Delta {f}(0)=
\frac{m^2 \gamma^2 b^2 k^2}{8 \pi^3 \hbar^4} P \int
\frac{q^4}{q^2-k^2} \frac{(4 \pi Z e)^2}{[(q^2- 2kq cos \vartheta
+ k^2)+\kappa^2]^2} sin ^3 \vartheta d \vartheta d q,
 \label{DeltaRef1}
\end{equation}
i.e.
\begin{equation}
 Re \Delta {f}(0)=
\frac{m^2 \gamma^2 b^2 k^2}{8 \pi^3 \hbar^4} P \int_0^{\infty}
\int_{-1}^{+1} \frac{q^4}{q^2-k^2}(4 \pi Z e)^2 \frac{1-x^2}{[q^2-
2kq x + k^2+\kappa^2]^2}  d x d q,
 \label{DeltaRef2}
\end{equation}
Integration over $x$ can be done explicitly that provides:
\begin{equation}
 Re \Delta {f}(0)=
\frac{m^2 \gamma^2 b^2 k^2}{8 \pi^3 \hbar^4}(4 \pi Z e)^2 P
\int_0^{q_max} \frac{q^4}{q^2-k^2} J(q) d q,
 \label{DeltaRef3}
\end{equation}
where
\begin{equation}
J(q)= - \frac{1}{[(q^2+ k^2+\kappa^2)^2 - 4k^2 q^2}
\left(
(\frac{q^2+k^2+\kappa^2}{kq}-2) - \frac{1}{2 k^2 q^2} - \frac{2
(q^2+k^2 + \kappa^2)}{4(kq)^3} ln \frac{q^2+k^2+\kappa^2
-2kq}{q^2+k^2+\kappa^2 + 2kq}
\right),
 \label{J}
\end{equation}

 The expression under the integration in (\ref{DeltaRef3})
 diverges on the upper limit as $q^2$. This growth is caused by
 the  point nature of interaction in the considered example.
 If consider the function $\rho(k-q)$, this growth disappears.

 \section{Conclusion}

Spin-orbital interaction contributes to the birefringence effect
for particles with the spin $S \ge 1$ along with central
interaction.
More detailed calculation of birefringence effect influence on the
process of production (photoproduction) of particles with the spin
$S \ge 1$ in nuclei will be done separately.


\end{document}